\documentclass[aps,prl,showpacs,preprint]{revtex4-1}
\usepackage{graphicx}
\begin{document}

\title{Imaging and quantum efficiency measurement of chromium emitters in diamond}
\author{I. Aharonovich$^{1}$, S. Castelletto$^{1}$, B. C. Gibson, B. C. Johnson, and S. Prawer}
\address{School of Physics, The University of Melbourne, 3010 Victoria, Australia}
\email{i.aharonovich@pgrad.unimelb.edu.au,sacas@unimelb.edu.au} 

\begin{abstract}
We present direct imaging of the emission pattern of individual chromium-based single photon emitters
in diamond and measure their quantum efficiency. By imaging the excited state transition dipole
intensity distribution in the back focal plane of high numerical aperture objective, we determined
that the emission dipole is oriented nearly orthogonal to the diamond-air interface. Employing
ion implantation techniques, the emitters were engineered with various proximities from the
diamond-air interface. By comparing the decay rates from the single chromium emitters at different
depths in the diamond crystal, an average quantum efficiency of 28\% was measured.
\end{abstract}

\maketitle

Quantum efficiency is a fundamental property of any nanoscopic
emitter since it dictates the ability to emit a photon once an excitation photon is absorbed.
The quantum efficiency (QE) is  defined as
QE$=k^{\infty}_{rad}/(k_{nr}+k^{\infty}_{rad})$, where $k^{\infty}_{rad}$ and $k_{nr}$ are the radiative decay rate in an homogenous unbounded medium and the non radiative decay rate of the emitter, respectively. A knowledge of the quantum efficiency is vital for applications requiring a single photon source on demand (optical quantum computing\cite{Obrien09} and quantum metrology, e.g. "quantum candela"\cite{Cheung07}) and for tackling challenges such as strong light-atom interaction or long distance entanglement protocols by means of integrated waveguides and microcavities in solid state systems\cite{Barclay09,Schietinger09}. Furthermore, to design optimal optical structures (e.g. nanocavities or plasmonic), which enhance the collection efficiency or modifies the radiative or/and non-radiative decay paths \cite{Kinkhabwala09, Muskens07, Taminiau08, Schietinger09, Barclay09,Vion10}, a precise measurement of QE is necessary to accurately quantify any improvement obtained in the detected photons from the coupling.

Direct measurements of QE are challenging and require a
priori knowledge of the emission dipole orientation, and a separate
measurement of the radiative and non radiative decay rates. To
determine the dipole orientation of single molecules and colloidal
quantum dots, methods such as defocused \cite{Bohmer03,Patra05},
direct imaging \cite{Novotny04}, or near field microscopy \cite{Betzig91,VanHulst00,Drezet04} have been successfully demonstrated.
This is achieved by imaging the emission pattern of the collected
photons with a high numerical aperture objective in the back focal
plane (or the back-aperture plane for an infinity corrected system).
\begin{figure}[htb]
\centerline{
\includegraphics[width=8.3cm]{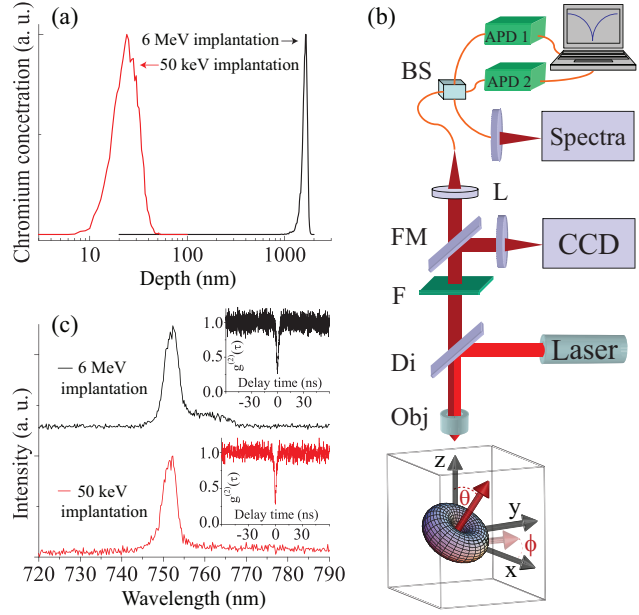}}
\caption{(a) Concentration profiles of implanted chromium ions into diamond using an acceleration voltage of 50 keV (red curve) and 6 MeV (black curve) determined using SRIM. (b) Schematic illustration of the confocal microscope (Obj 100x 0.95 numerical aperture objective; Di dichroic mirror; F bandpass filter, FM flip mirror; BS beam splitter, APD$_{1,2}$ single photon detectors), with a CCD imaging channel. An illustration of the emitting
dipole orientation with respect to the diamond sample and the
optical axis (z) of the objective, identified by the polar angle $\theta$ and the azimuth angle $\Phi$. (c) Example of PL spectra of chromium related centers with the same zero phonon line created by ion implantation using energies of 50 keV (red curve) and a 6 Mev (black curve). Insets show antibunching curves demonstrating single photon emission.}\label{SRIM}
\end{figure}
To extract information about radiative and non-radiative decay
rates for an emitter close to the material-air interface, one can
modify the local dielectric environment of the emitters. This is enabled by adding to the emitter environment a medium
with a matching \cite{Brokmann04} or a different refractive
index \cite{Macklin96}. Using such approaches, radiative and non radiative decay rates of single molecules and quantum
dots were separately obtained. In an alternative method, a
scanning metal mirror was brought close to a single molecular
dipole. From the modification of the molecule radiative decay rate,
the QE of single emitting dipole was then measured \cite{Buchler05}.

Recently, high brightness single photon emitters
originating from chromium impurities in diamond were fabricated
\cite{Aharonovich09,Aharonovich10b}. The optical properties of these
centers reveal some outstanding features compared to other centers
in diamond, such as very narrow spectral emission
at room temperature (a few nanometers), short excited state life
time of $<$ 3.5 ns and large dipole moment. These attributes make them leading candidates for applications in quantum information science\cite{Obrien09}, sub-diffraction microscopy\cite{Rittweger09} and biological systems\cite{Fu07}.

In this letter we implement ion implantation and imaging techniques to measure directly the quantum efficiency of single centers in
monolithic diamond, demonstrated with chromium impurities. For this purpose the orientation of the emitter has to be known and the decay rate of the emitter has to be measured in two different dielectric environments. Employing this approach, we first imaged the
emission patterns of single chromium emitters to identify
their emission dipole orientation. We then measured the decay rates
from emitters located in close proximity to the diamond surface
and from emitters located deep in the diamond crystal. The information obtained enabled the determination of the
 quantum efficiency of individual single photon emitters.

To fabricate the emitters close to the diamond-air interface, chromium ions (fluence of 10$^{10}$ ions/cm$^2$)  and oxygen ions (fluence of 1.5 $\times$ 10$^{10}$ ions/cm$^2$) were accelerated to 50 keV and 19.5 keV, respectively and implanted into a (100) oriented type IIA diamond ([N]< 1ppm, [B]< 0.05 ppm). To modify the dielectric environment of the emitters, chromium and oxygen were implanted into the same type of diamond using the same fluencies and an acceleration voltage of 6 MeV and 3.6 MeV, respectively.

The samples were then
annealed to 1000 $^{\circ}$C for two hours under a forming gas
(95\%Ar-5\%H$_2$) ambient. Fig.(\ref{SRIM})a shows the simulation
of the implantations using stopping range of ions in matter (SRIM).
From the simulation it is evident that the projected range of the
shallow implantation (50 keV) is approximately 25 nm below the diamond
surface, while the projected range of the deep implantation (6 Mev) is 1.5
$\mu$m below the diamond surface. Note that the annealing step
applied after the implantation is not sufficient to cause any
diffusion of the Cr atoms and the end of range of the two implanted
chromium ions does not overlap.

The samples were scanned using a confocal microscope as depicted
schematically in Fig.(\ref{SRIM})b. Single photon emitters were
first identified using an Hanbury-Brown and Twiss (HBT)
interferometer and their photoluminescence (PL) spectra were
recorded, as shown in Fig.(\ref{SRIM})c. The typical PL emission occurs in the range of 748-760 nm.
Fig.(\ref{SRIM})c clearly demonstrates that emitters with the same zero phonon line (ZPL) can be fabricated by either shallow or deep ion implantation. A second detection channel was added after the dichroic mirror in the confocal setup to image the
transition dipole of individual single photon emitters by an imaging
lens and a cooled CCD camera with quantum efficiency of 40\% at 750 nm.

In the first part of the experiment we imaged the emission dipole orientation by recording the
angular intensity distribution of single emitters in the back focal
plane of a high numerical aperture objective using a CCD camera.
Such images of single emitters are crucial as they provide a clear
indication regarding the dipole orientation.
\begin{figure}[htb]
\centerline{
\includegraphics[width=7.3cm]{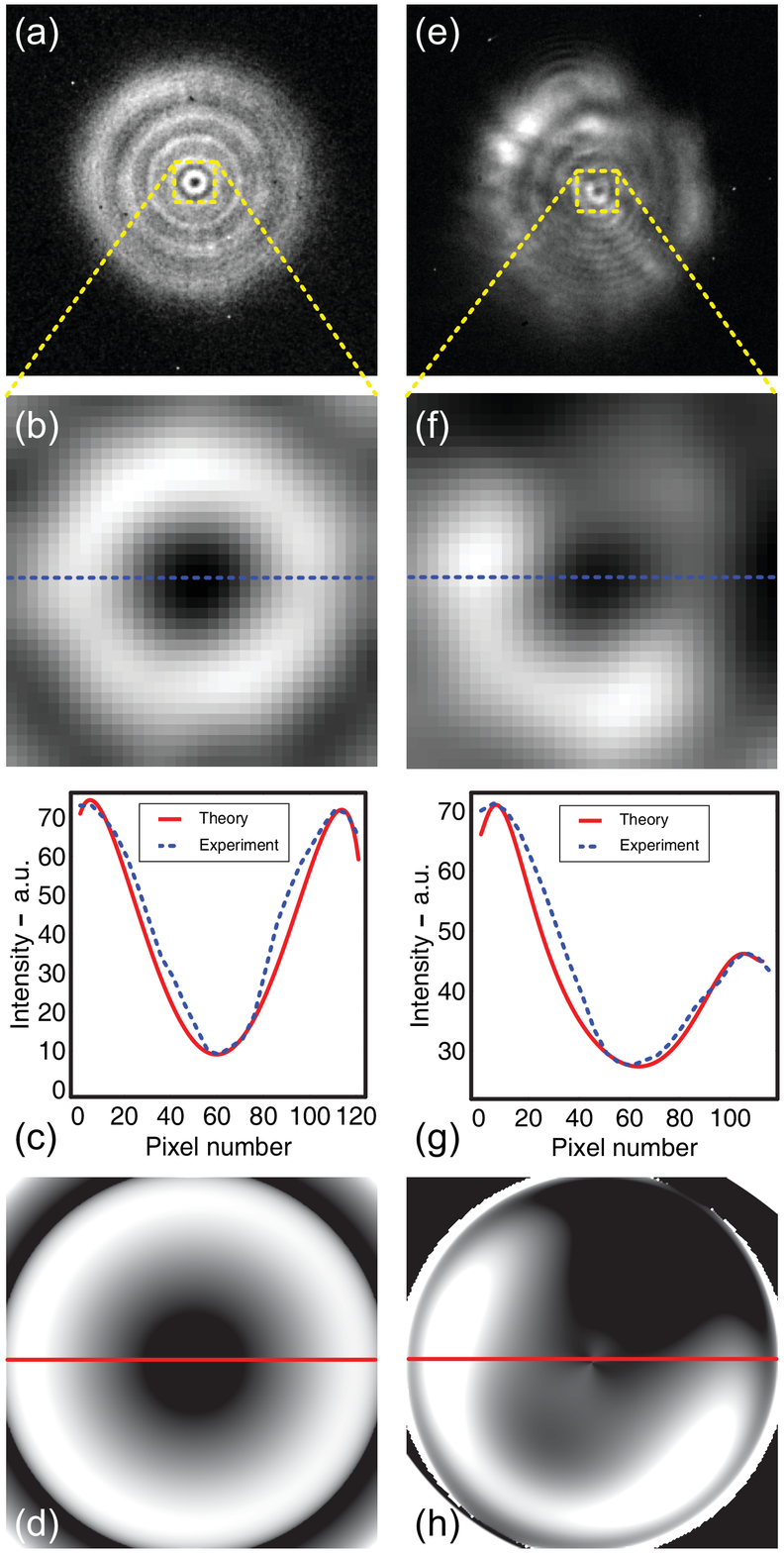}}
\caption{(a,e) Two images of the intensity distribution
of the emission dipole, from chromium single photon emitters (ZPL at 753 nm) created in bulk diamond and in sub-micron diamond crystal, respectively. Integration times were 200 s  and 60 s, respectively. (b,f) Magnified area of the central
ring of the images depicted in (a,e). (c,g) The cross section experimental data
and the theoretical fit of the emission pattern are shown in (b,f).
 The polar angles of the emitters are $\theta=(1\pm 1)^{\circ}$ and $\theta=(49 \pm 2)^{\circ}$ for the bulk and sub-micron diamond, respectively, while the azimuth angles are $\phi=(0\pm 5)^{\circ}$ and $\phi=(69\pm 2)^{\circ}$.(d, h) Calculated pattern of the dipole emission shown in (b,f)
 using the parameters from the fit. An excellent agreement between the theory and the experimental data is obtained
 for each dipole orientation.}
\label{CCD}
\end{figure}
Fig.
(\ref{CCD})(a) shows a typical objective back-focal-plane dipole image recorded from a single chromium center in bulk diamond, that differs significantly from a standard Airy point spread function. Concentric doughnut-shaped rings, associated with being imaged through the aperture of a dry objective, are observed in
the CCD image \cite{Dickson98}.
 The uniform intensity distribution of the bright rings with dark centers indicates that the emitter is oriented nearly orthogonal to the diamond-air interface. This confirms that the Cr center is not aligned along a <111> direction neighboring a vacancy, as then a trigonal symmetry is expected. More than 20 single emitters were imaged individually, all confirming a very similar dipole orientations, with polar angles between 0 and 2 degrees, which are within our method sensitivity.

For the sake of comparison, Fig.(\ref{CCD})(e) shows a typical objective back-focal plane dipole image of the Cr centers created in sub-micron diamond (300 nm average size), with the same ZPL as the emitter shown in Fig.(\ref{CCD}(a). In this case, as expected, the emission dipole orientation changes from crystal to crystal and it is clearly not parallel to the optics axis.  Fig.(\ref{CCD})(b,f) show a magnified area of the central rings of the images shown in (a,e), respectively.
Fig.(\ref{CCD})(c,g) show the cross section data and the fit of
the emission pattern shown in Fig.(\ref{CCD})(b,f), respectively,
according to the theory presented in \cite{Novotny04}. From the fit,
the dipole polar coordinate, $\theta$, was estimated to be less than (1$\pm 1)^\circ$ and the azimuth angle $\Phi=(0 \pm 5)^\circ$ for the bulk diamond. While for this particular nanocrystal, $\theta$=(49$\pm 2)^\circ$ and $\Phi=(69 \pm 2)^\circ$.
Figure(\ref{CCD})(d,h) show a two dimensional calculated pattern of
the dipole emission shown in Fig.(\ref{CCD})(b,f) using the
parameters from the fit. Excellent agreement between the theory
and experiment is obtained for the dipole orientation measurement.
Dipole imaging technique can be successfully applied to color centers in bulk and nano-diamonds to fully determine their 3D orientation.

In our previous work \cite{Aharonovich10b}, it was determined that
the absorption dipole for Cr centres in bulk diamond is aligned along
one of the main crystallographic axis on the plane of the surface and the emitted light is not
linearly polarized. The nearly orthogonal emission dipole observed in this work elucidates that the
emission dipole of the chromium centers in bulk diamond is nearly perpendicular to its absorption dipole.

In the second part of our experiment, we measured the total excited
state lifetime and the QE of individual emitters in bulk diamond.
It is well known that the radiative lifetime of an emitter in a
homogeneous medium of refractive index $n$ is inversely proportional
to $n$. In the more complex situation of a linear dipole located at
a distance $d < \lambda$ from a dielectric interface, Lukosz and
Kunz \cite{Kunz77} showed that the radiative decay rate ($k_{rad}$)
depends on the distance $d$, the refractive index of each dielectric
medium and the excitation dipole orientation polar angle $\theta$,
with respect to the normal to the interface.

We denote by $k_{\infty}=
k_{nr}+k^{\infty}_{rad}$, the total decay rate of an
emitter in an unbounded homogeneous medium and
$\alpha(d,\theta, n_1)=k_{rad}(d,\theta,n_1)/k_{rad}^\infty$ the
modification of the decay rate in the presence of the dielectric
interface, where $n_1=n_2/n$, being $n_2$ the index of refraction of
the medium after the interface.
 The physical interpretation can be
qualitatively described by classical
 electrodynamics. When the dipole radiates, its field is partly reflected by the interface.
 The dipole can then interact with its own field. This self interaction modifies the oscillation amplitude
 (and frequency) of the dipole and, as a consequence, affects its radiative decay time.
The total decay rate for a linear dipole can thus be generally written as
\begin{equation}
 k(d, \theta)= k_{nr}+k^{\infty}_{rad} [\alpha(d, n_1)_{\parallel} sin^2(\theta)+\alpha(d, n_1)_{\perp} cos^2(\theta)]
 \label{total decay rate}
\end{equation}
where $\alpha_{{\parallel},_{\perp}}$ refers to a parallel and
orthogonal dipole to the interface, with the algebraic expression given
in \cite{Kunz77}. If an emitter is moved far from the interface or the refractive index
difference of the interface is reduced to zero both
$\alpha_{{\parallel},_{\perp}}$=1, and the excited state lifetime is
independent of the dipole orientation.
From
the decay rates for dipoles close to an interface and in an unbounded medium, we deduce the value $\beta=k(d, \theta_e)/k_{\infty}$.
The QE can thus be rewritten as
QE$=(1-\beta)/[1-\alpha(d,\theta,n_1)]$\cite{Brokmann05}.

 In our particular case the
chromium centers implanted with energies of 6 MeV are considered to be in an unbounded medium ($d\simeq$ 1.5 $\mu$m $>\lambda$) and far from the diamond-air boundary, while centers created using a 50 keV implantation are located near a dielectric interface. Therefore, measuring the excited
state lifetime of deep implants will provide direct information of $k_{\infty}$, while measuring the decay rates of chromium centers engineered near the surface will allow to deduce $k(d, \theta)$. To exclude any wavelength dependent effect, only emitters with the same peak emission are compared.  Note that since the centers are embedded in the diamond
matrix, the immediate surroundings in both the shallow and the deep
implantations are the same and therefore $k_{nr}$ can be assumed to
remain constant \cite{Brokmann05}.

Employing a pulsed laser at a wavelength of 690 nm with pulse width of 20 ps and
repetition rate of 40 MHz, we measured directly the total excited
state lifetime of various emitters in the shallow and deep
implantation configuration (implanted into a bulk single crystal diamond).
\begin{figure}[htb]
\centerline{
\includegraphics[width=8.3cm]{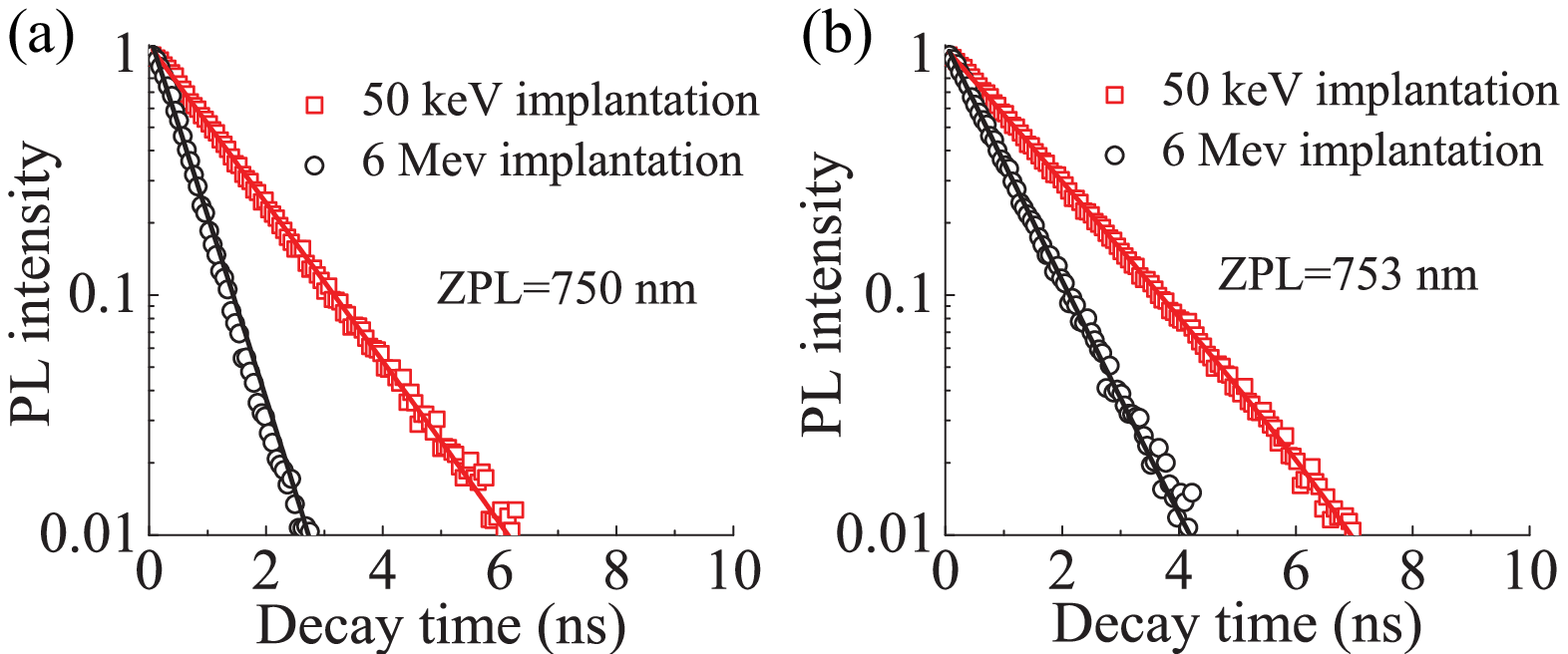}}
\caption{(a) Direct lifetime measurement of a single emitter with a
ZPL centered at 750 nm close to the diamond-air interface (squares)
and of an emitter with the same ZPL located 1.5 $\mu$m below the
diamond surface (circles). The data were fit using a single
exponential fit (solid line). (b) Decay rate measurements recorded
from a different emitter with a ZPL centered at 753 nm located near the interface (squares) and deep in the diamond crystal (circles)}
\label{decay}
\end{figure}
Fig.(\ref{decay})a,b show the fluorescence decay rates of two single chromium emitters for both
shallow and deep implantations. The angle $\theta$ of the emission dipole of these emitters was found to be $\theta=(1 \pm 1)^{\circ}$ (Fig.\ref{CCD}a) and $\theta=(0.5 \pm 1)^{\circ}$ (not shown here), respectively. The data were fitted by using
mono-exponential curve with a relative uncertainty of 0.5\%. The
reduction of the total decay rate for emitters located near the
diamond air interface is clearly seen from these measurements.
From
the measured decay rates for the shallow and the deep
implantations, the value $\beta=k(d, \theta)/k_{\infty}$ was deduced, as an average obtained for emitters with the same ZPL.
For two single emitters at 750 nm and 753 nm, the calculated
parameter $\alpha$, and the measured value of $\beta$, yield a
QE=0.42$\pm$0.06 and QE=0.29$\pm$0.05, respectively. To the best of our
knowledge, this is the first direct measurement of QE of a single color center in diamond.

In the last part of our experiment, we measured the average QE of an ensemble of single chromium emitters in bulk diamond, regardless of
their peak emission wavelength.
The average excited state
lifetime for the centers located near the diamond-air interface is 1.24$\pm$0.13
ns, while the lifetime of the centers located deep in the diamond
crystal is 0.92$\pm$0.09 ns. A clear reduction of the excited state
lifetime for the emitters located in an unbounded medium is
noticeable also in an ensemble measurement. This result confirms
that the centers are associated with a linear dipole since a 2D
dipole orthogonal to the surface would not provide such a variation
in the measured excited state life time in the shallow and deep implantation \cite{Brokmann05}.

\begin{figure}[htb]
\centerline{
\includegraphics[width=8.3cm]{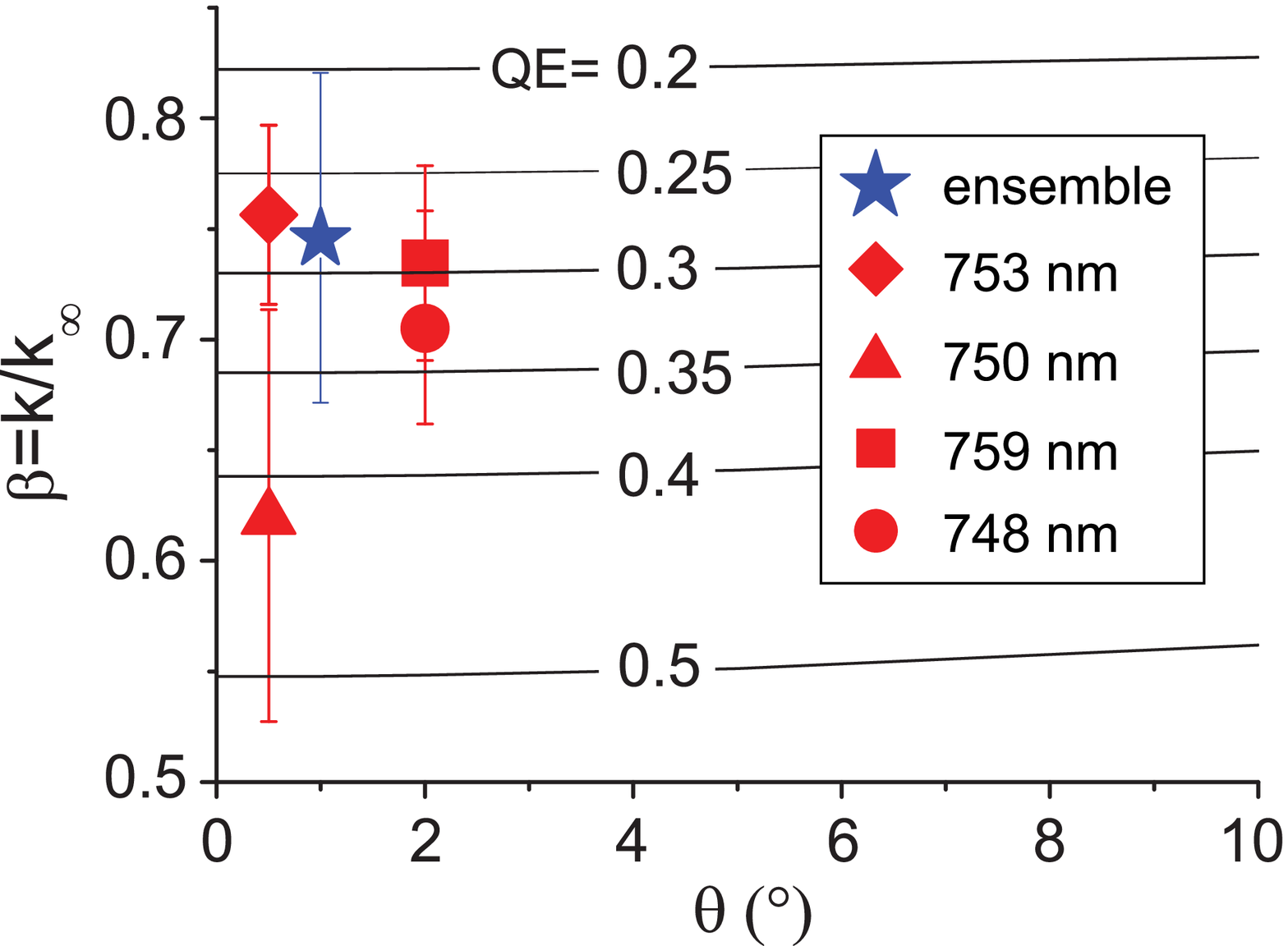}}
\caption{Orientational dependence of $\beta$ on the polar angle. The solid black lines correspond to
 calculated values for different
values of QE and with $d=25\pm5$nm. The measured values of $\beta$ from an
ensemble measurement (blue star) and several single centers with different ZPL (red circle, square,
triangle and diamond) are superimposed.} \label{dist}
\end{figure}

The QE was computed for various polar angles and values of $\beta$ and is plotted in Fig.(\ref{dist})(black lines). The nearly orthogonal emission dipole observed for the centers yields
an ensemble value of $\alpha$ =0.098 \cite{Kunz77} and results in an averaged QE=0.28$\pm$0.04 for chromium emitters, regardless their peak emission wavelength. The experimental values of the QE of several single centers with the same ZPL and of the ensemble measurement are shown in Fig.(\ref{dist}).
A QE in the range of 30\% can be associated to the presence of a metastable state or to additional non radiative process such as decay through phonons, ionization or heat, which strongly depend on the environment. The inter-system crossing rate for the chromium emitters is $k_{ISC}$=5.1 MHz, as was deduced from analyzing the second order autocorrelation function. This value indicates that the population of the metastable state does not significantly reduce the value of the QE since the radiative decay is $\sim$ 313 MHz and the non-radiative decay $\sim$ 769 MHz.

A number of important implications can be drawn from these results.
The peculiarity of an emission dipole nearly always orthogonal to
the bulk diamond surface is particularly advantageous for the integration
of these emitters with cavities or diamond nano-antennas. In fact
the major drawback of the recently fabricated diamond antennas
incorporating NV$^{-}$ centers \cite{Babinec10} was the nondeterministic
emission dipole orientation of the NV$^{-}$ centers, due to its trigonal
symmetry and polarization absorption anisotropy. This drawback can
be overcome by using the chromium emitters in similar geometries and
by using bulk diamond crystals with different crystallographic
orientations.
Finally, a similar approach could be used to determine the actual QE of NV$^{-}$, which is commonly inferred on the basis of circumstantial evidence \cite{Collins83,Jelezko06}.

To summarize, we present for the first time, the emission dipole
pattern images of single color centers in bulk and nanodiamonds and a direct
measurement of their quantum efficiency in bulk diamond. The dipoles are nearly
orthogonal to the bulk diamond-air interface and to its absorption
dipole.
Finally, by employing ion implantation techniques, we were able to
fabricate the emitters at various distances from the diamond
surface, thus modifying their radiative lifetime. Combining the
imaging of the dipoles and measuring the decay rates of the emitters
close to the diamond-air interface and in the unbounded medium, the
quantum efficiency of individual centers and of an ensemble of centers in monolithic diamond
was determined.
\\

We thank A. Tierney for useful discussions. The Department of Electronic Materials Engineering at the Australian
National University is acknowledged for their support by providing access to
ion implanting facilities. This work was supported The International Science
Linkages Program of the Australian Department of Innovation,
Industry, Science and Research (Project No. CG110039), the Australian Research Council and by the
European Union Sixth Framework Program under the EQUIND IST-034368.
\\

$^{1}$ IA and SC equally contributed to the paper.

\end{document}